\documentclass{article}
\usepackage{spconf,amsmath,graphicx}
\usepackage[colorlinks]{hyperref}
\usepackage{url}
\usepackage{amssymb,amsfonts,bm}
\usepackage{multirow}

\graphicspath{{figs/}}

\newcommand{\bfsl}[1]{\textbf{\textsl{#1}}}

\title{CycleGAN-VC2:\\Improved CycleGAN-based Non-parallel Voice Conversion}
\name{Takuhiro Kaneko, Hirokazu Kameoka, Kou Tanaka, Nobukatsu Hojo}
\address{NTT Communication Science Laboratories, NTT Corporation, Japan}

\begin{document}
\maketitle
\begin{abstract}
  Non-parallel voice conversion (VC) is a technique for learning the mapping from source to target speech without relying on parallel data. This is an important task, but it has been challenging due to the disadvantages of the training conditions. Recently, CycleGAN-VC has provided a breakthrough and performed comparably to a parallel VC method without relying on any extra data, modules, or time alignment procedures. However, there is still a large gap between the real target and converted speech, and bridging this gap remains a challenge. To reduce this gap, we propose CycleGAN-VC2, which is an improved version of CycleGAN-VC incorporating three new techniques: an improved objective (two-step adversarial losses), improved generator (2-1-2D CNN), and improved discriminator (PatchGAN). We evaluated our method on a non-parallel VC task and analyzed the effect of each technique in detail. An objective evaluation showed that these techniques help bring the converted feature sequence closer to the target in terms of both global and local structures, which we assess by using Mel-cepstral distortion and modulation spectra distance, respectively. A subjective evaluation showed that CycleGAN-VC2 outperforms CycleGAN-VC in terms of naturalness and similarity for every speaker pair, including intra-gender and inter-gender pairs.\footnote{The converted speech samples are provided at \url{http://www.kecl.ntt.co.jp/people/kaneko.takuhiro/projects/cyclegan-vc2/index.html}.}
\end{abstract}


\section{Introduction}
\label{sec:intro}
Voice conversion (VC) is a technique for transforming the non/para-linguistic information of given speech while preserving the linguistic information. VC has great potential for application to various tasks, such as speaking aids~\cite{AKainSC2007,KNakamuraSC2012} and the conversion of style~\cite{ZInanogluSC2009,TTodaTASLP2012} and pronunciation~\cite{TKanekoIS2017b}.

One successful approach to VC involves statistical methods based on Gaussian mixture model (GMM)~\cite{YStylianouTASP1998,TTodaTASLP2007,EHelanderTASLP2010}, neural network (NN)-based methods using restricted Boltzmann machines (RBMs)~\cite{LChenTASLP2014,TNakashikaIEICE2014}, feed forward NNs (FNNs)~\cite{SDesaiTASLP2010,SMohammadiSLT2014,OKeisukeAPSIPAASC2017}, recurrent NNs (RNNs)~\cite{TNakashikaIS2014,LSunICASSP2015}, convolutional NNs (CNNs)~\cite{TKanekoIS2017b}, attention networks~\cite{KTanakaICASSP2019,HKameokaArXiv2018b}, and generative adversarial networks (GANs)~\cite{TKanekoIS2017b}, and exemplar-based methods using non-negative matrix factorization (NMF)~\cite{RTakashimaIEICE2013,ZWuTASLP2014}.

Many VC methods (including the above-mentioned) are categorized as parallel VC, which relies on the availability of parallel utterance pairs of the source and target speakers. However, collecting such data is often laborious or impractical. Even if obtaining such data is feasible, many VC methods require a time alignment procedure as a pre-process, which may occasionally fail and requires careful pre-screening or manual correction. To overcome these restrictions, this paper focuses on non-parallel VC, which does not rely on parallel utterances, transcriptions, or time alignment procedures.

In general, non-parallel VC is quite challenging and is inferior to parallel VC in terms of quality due to the disadvantages of the training conditions. To alleviate these severe conditions, several studies have incorporated an extra module (e.g., an automatic speech recognition (ASR) module~\cite{FLXieIS2016,YSaitoICASSP2018}) or extra data (e.g., parallel utterance pairs among reference speakers~\cite{AMouchtarisTASLP2006,CHLeeIS2006,TTodaIS2006,DSaitoIS2011}). Although these additional modules or data are helpful for training, preparing them imposes other costs and thus limits application. To avoid such additional costs, recent studies have examined the use of probabilistic NNs (e.g., an RBN~\cite{TNakashikaTASLP2016} and variational autoencoders (VAEs)~\cite{CHsuIS2017,HKameokaArXiv2018}), which embed the acoustic features into common low-dimensional space with the supervision of speaker identification. It is noteworthy that they are free from extra data, modules, and time alignment procedures. However, one limitation is that they need to approximate data distribution explicitly (e.g., Gaussian is typically used), which tends to cause over-smoothing through statistical averaging.

To overcome these limitations, recent studies~\cite{CHsuIS2017,TKanekoArXiv2017,FFangICASSP2018} have incorporated GANs~\cite{IGoodfellowNIPS2014}, which can learn a generative distribution close to the target without explicit approximation, thus avoiding the over-smoothing caused by statistical averaging. Among these, in contrast to some of the frame-by-frame methods~\cite{CHsuIS2017,FFangICASSP2018}, which have difficulty in learning time dependencies, CycleGAN-VC~\cite{TKanekoArXiv2017} (published in \cite{TKanekoEUSIPCO2018}) makes it possible to learn a sequence-based mapping function by using CycleGAN~\cite{JYZhuICCV2017,ZYiICCV2017,TKimICML2017} with a gated CNN~\cite{YDauphinICML2017} and identity-mapping loss~\cite{YTaigmanICLR2017}. This allows sequential and hierarchical structures to be captured while preserving linguistic information. With this improvement, CycleGAN-VC performed comparably to a parallel VC method~\cite{TTodaTASLP2007}.

However, even using CycleGAN-VC, there is still a challenging gap to bridge between the real target and converted speech. To reduce this gap, we propose CycleGAN-VC2, which is an improved version of CycleGAN-VC incorporating three new techniques: an improved objective (two-step adversarial losses), improved generator (2-1-2D CNN), and improved discriminator (PatchGAN). We analyzed the effect of each technique on the Spoke (i.e., non-parallel VC) task of the Voice Conversion Challenge 2018 (VCC 2018)~\cite{VCC2018}. An objective evaluation showed that the proposed techniques help bring the converted acoustic feature sequence closer to the target in terms of global and local structures, which we assess by using Mel-cepstral distortion and modulation spectra distance, respectively. A subjective evaluation showed that CycleGAN-VC2 outperforms CycleGAN-VC in terms of naturalness and similarity for every speaker pair, including intra-gender and inter-gender pairs.

In Section~\ref{sec:cyclegan-vc} of this paper, we review the conventional CycleGAN-VC. In Section~\ref{sec:cyclegan-vc2}, we describe CycleGAN-VC2, which is an improved version of CycleGAN-VC incorporating three new techniques. In Section~\ref{sec:exp}, we report the experimental results. We conclude in Section~\ref{sec:conc} with a brief summary and mention future work.

\section{Conventional CycleGAN-VC}
\label{sec:cyclegan-vc}
\vspace{-1mm}
\subsection{Objective: One-Step Adversarial Loss}
\label{subsec:obj}
Let $x \in \mathbb{R}^{Q \times T_x}$ and $y \in \mathbb{R}^{Q \times T_y}$ be acoustic feature sequences belonging to source $X$ and target $Y$, respectively, where $Q$ is the feature dimension and $T_x$ and $T_y$ are the sequence lengths. The goal of CycleGAN-VC is to learn mapping $G_{X \rightarrow Y}$, which converts $x \in X$ into $y \in Y$, without relying on parallel data. Inspired by CycleGAN~\cite{JYZhuICCV2017}, which was originally proposed in computer vision for unpaired image-to-image translation, CycleGAN-VC uses an adversarial loss~\cite{IGoodfellowNIPS2014} and cycle-consistency loss~\cite{TZhouCVPR2016}. Additionally, to encourage the preservation of linguistic information, CycleGAN-VC also uses an identity-mapping loss~\cite{YTaigmanICLR2017}.

{\bf Adversarial loss:}
To make a converted feature $G_{X \rightarrow Y}(x)$ indistinguishable from a target $y$, an adversarial loss is used:
\begin{flalign}
  \label{eqn:adv}
  &{\cal L}_{adv} (G_{X \rightarrow Y}, D_Y)
  = \mathbb{E}_{y \sim P_Y(y)} [ \log D_Y(y) ]
  \nonumber \\
  &\:\:\:\:\:\:\:\:\:\:\:\:\:\:\:\:\:\:\:\:\:\:\:\:\:\:
  + \mathbb{E}_{x \sim P_X(x)} [ \log (1 - D_Y(G_{X \rightarrow Y}(x))) ],
\end{flalign}
where discriminator $D_Y$ attempts to find the best decision boundary between real and converted features by maximizing this loss, and $G_{X \rightarrow Y}$ attempts to generate a feature that can deceive $D_Y$ by minimizing this loss.

{\bf Cycle-consistency loss:}
The adversarial loss only restricts $G_{X \rightarrow Y}(x)$ to follow the target distribution and does not guarantee the linguistic consistency between input and output features. To further regularize the mapping, a cycle-consistency loss is used:
\begin{flalign}
  \label{eqn:cycle}
  &\: {\cal L}_{cyc}(G_{X \rightarrow Y}, G_{Y \rightarrow X})
  \nonumber \\
  = &\: \mathbb{E}_{x \sim P_X(x)} [ \|
  G_{Y \rightarrow X}( G_{X \rightarrow Y}(x) ) - x \|_1 ]
  \nonumber \\
  + &\: \mathbb{E}_{y \sim P_Y(y)} [ \|
  G_{X \rightarrow Y}( G_{Y \rightarrow X}(y) ) - y \|_1 ],
\end{flalign}
where forward-inverse and inverse-forward mappings are simultaneously learned to stabilize training. This loss encourages $G_{X \rightarrow Y}$ and $G_{Y \rightarrow X}$ to find an optimal pseudo pair of $(x, y)$ through circular conversion, as shown in Fig.~\ref{fig:objective}(a).

{\bf Identity-mapping loss:}
To further encourage the input preservation, an identity-mapping loss is used:
\begin{flalign}
  \label{eqn:identity}
  &{\cal L}_{id}(G_{X \rightarrow Y}, G_{Y \rightarrow X})
  = \mathbb{E}_{y \sim P_Y(y)} [ \|
  G_{X \rightarrow Y}(y) - y \|_1]
  \nonumber \\
  &\:\:\:\:\:\:\:\:\:\:\:\:\:\:\:\:\:\:\:\:\:\:\:\:\:\:
  + \mathbb{E}_{x \sim P_X(x)} [ \|
  G_{Y \rightarrow X}(x) - x \|_1].
\end{flalign}

{\bf Full objective:}
The full objective is written as
\begin{flalign}
  \label{eqn:full}
  &{\cal L}_{full} =
  {\cal L}_{adv} (G_{X \rightarrow Y}, D_Y)
  + {\cal L}_{adv} (G_{Y \rightarrow X}, D_X)
  \nonumber \\
  &\: + \lambda_{cyc} {\cal L}_{cyc}(G_{X \rightarrow Y}, G_{Y \rightarrow X})
  + \lambda_{id} {\cal L}_{id}(G_{X \rightarrow Y}, G_{Y \rightarrow X}),
\end{flalign}
where $\lambda_{cyc}$ and $\lambda_{id}$ are trade-off parameters. In this formulation, an adversarial loss is used once for each cycle, as shown in Fig.~\ref{fig:objective}(a). Hence, we call it a \textit{one-step adversarial loss}.

\subsection{Generator: 1D CNN}
\label{subsec:g}
CycleGAN-VC uses a \textit{one-dimensional (1D) CNN}~\cite{TKanekoIS2017b} for the generator to capture the overall relationship along with the feature direction while preserving the temporal structure. This can be viewed as the direct temporal extension of a frame-by-frame model that captures such features' relationship only per frame. To capture the wide-range temporal structure efficiently while preserving the input structure, the generator is composed of downsampling, residual~\cite{KHeCVPR2016}, and upsampling layers, as shown in Fig.~\ref{fig:generator}(a). The other notable point is that CycleGAN-VC uses a gated CNN~\cite{YDauphinICML2017} to capture the sequential and hierarchical structures of acoustic features.

\subsection{Discriminator: FullGAN}
\label{subsec:d}
CycleGAN-VC uses a 2D CNN~\cite{TKanekoIS2017b} for the discriminator to focus on a 2D structure (i.e., 2D spectral texture~\cite{TKanekoICASSP2017}). More precisely, as shown in Fig.~\ref{fig:discriminator}(a), a fully connected layer is used at the last layer to determine the realness considering the input's overall structure. Such a model is called \textit{FullGAN}.

\section{CycleGAN-VC2}
\label{sec:cyclegan-vc2}
\begin{figure}[t]
  \centerline{\includegraphics[width=\columnwidth]{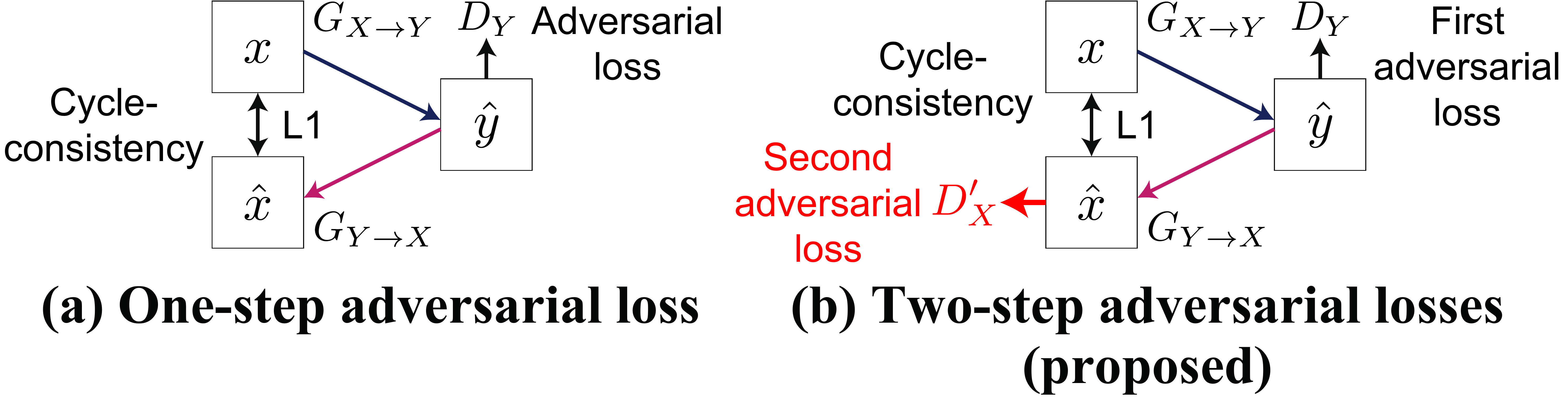}}
  \vspace{-3mm}
  \caption{Comparison of objectives}
  \vspace{-3mm}
  \label{fig:objective}
\end{figure}

\subsection{Improved Objective: Two-Step Adversarial Losses}
\label{subsec:imp_obj}
One well-known problem for statistical models is the over-smoothing caused by statistical averaging. The adversarial loss used in Eq.~\ref{eqn:full} helps to alleviate this degradation, but the cycle-consistency loss formulated as L1 still causes over-smoothing. To mitigate this negative effect, we introduce an additional discriminator $D'_X$ and impose an adversarial loss on the circularly converted feature, as
\begin{flalign}
  \label{eqn:adv2}
  {\cal L}_{adv2}(G_{X \rightarrow Y}, G_{Y \rightarrow X}, D'_X)
  = \mathbb{E}_{x \sim P_X(x)} [
  \log D'_X(x) ]
  \nonumber \\
  + \mathbb{E}_{x \sim P_X(x)} [
  \log (1 - D'_X( G_{Y \rightarrow X} (G_{X \rightarrow Y} (x)) )) ].
\end{flalign}
Similarly, we introduce $D'_Y$ and impose an adversarial loss ${\cal L}_{adv2}(G_{Y \rightarrow X}, G_{X \rightarrow Y}, D'_Y)$ for the inverse-forward mapping. We add these two adversarial losses to Eq.~\ref{eqn:full}. In this improved objective, we use adversarial losses twice for each cycle, as shown in Fig.~\ref{fig:objective}(b). Hence, we call them \textit{two-step adversarial losses}.

\begin{figure}[t]
  \centerline{\includegraphics[width=\columnwidth]{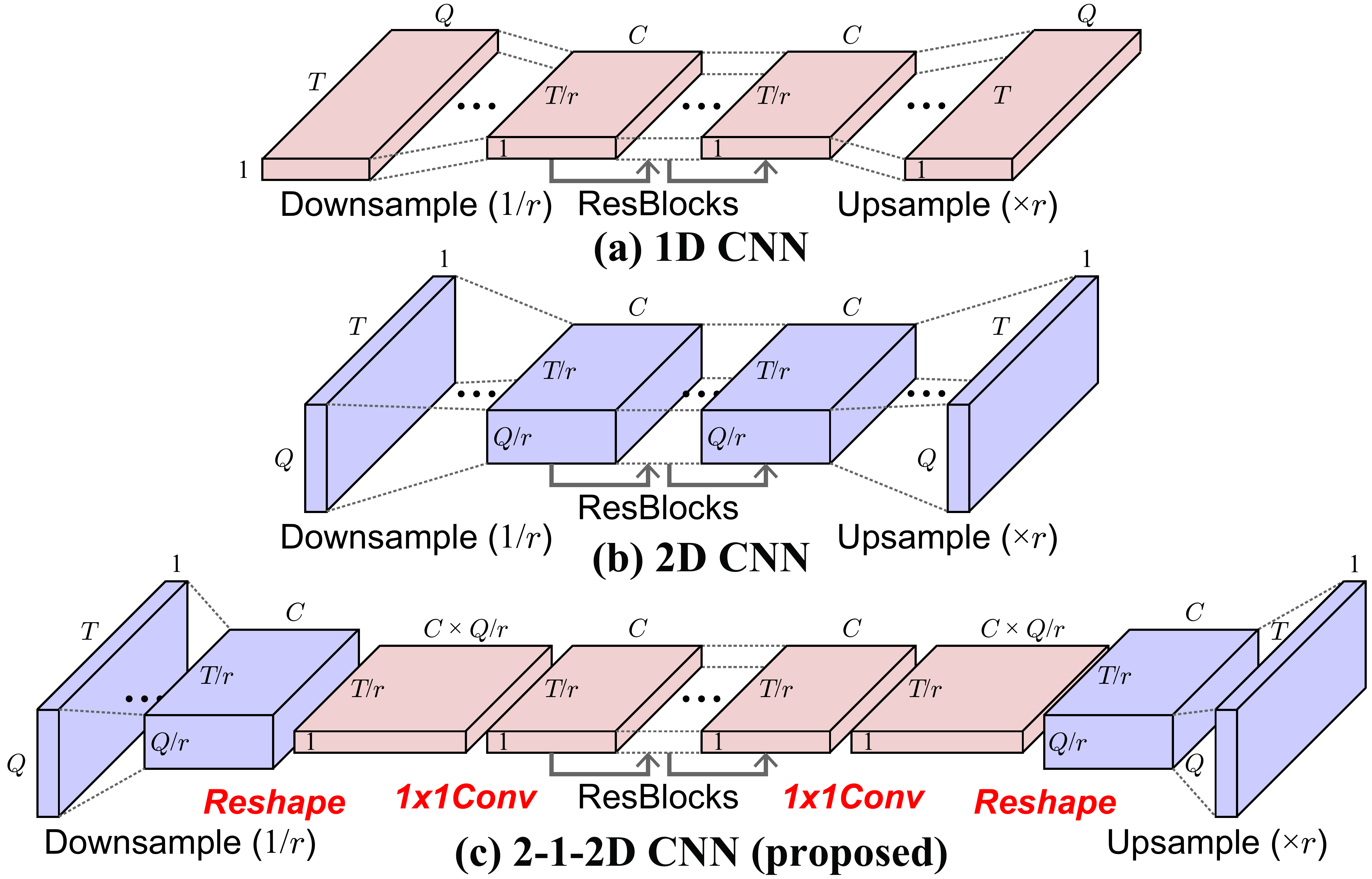}}
  \vspace{-3.5mm}
  \caption{Comparison of generator network architectures.
    Red and blue blocks indicate 1D and 2D convolution layers, respectively.
    $r$ indicates a downsampling or upsampling rate.}
  \vspace{-3mm}
  \label{fig:generator}
\end{figure}

\subsection{Improved Generator: 2-1-2D CNN}
\label{subsec:imp_g}
In a VC framework~\cite{TKanekoIS2017b,TKanekoArXiv2017} (including CycleGAN-VC), a 1D CNN (Fig.~\ref{fig:generator}(a)) is commonly used as a generator, whereas in a postfilter framework~\cite{TKanekoICASSP2017,TKanekoIS2017a}, a 2D CNN (Fig.~\ref{fig:generator}(b)) is more preferred. These choices are related to the pros and cons of each network. A 1D CNN is more feasible for capturing dynamical change, as it can capture the overall relationship along with the feature dimension. In contrast, a 2D CNN is better suited for converting features while preserving the original structures, as it restricts the converted region to local. Even using a 1D CNN, residual blocks~\cite{KHeCVPR2016} can mitigate the loss of the original structure, but we find that downsampling and upsampling (which are necessary for effectively capturing the wide-range structures) become a severe cause of this degradation. To alleviate it, we have developed a network architecture called a \textit{2-1-2D CNN}, shown in Fig.~\ref{fig:generator}(c). In this network, 2D convolution is used for downsampling and upsampling, and 1D convolution is used for the main conversion process (i.e., residual blocks). To adjust the channel dimension, we apply $1\times1$ convolution before or after reshaping the feature map.

\subsection{Improved Discriminator: PatchGAN}
\label{subsec:imp_d}
\begin{figure}[t]
  \centerline{\includegraphics[width=0.8\columnwidth]{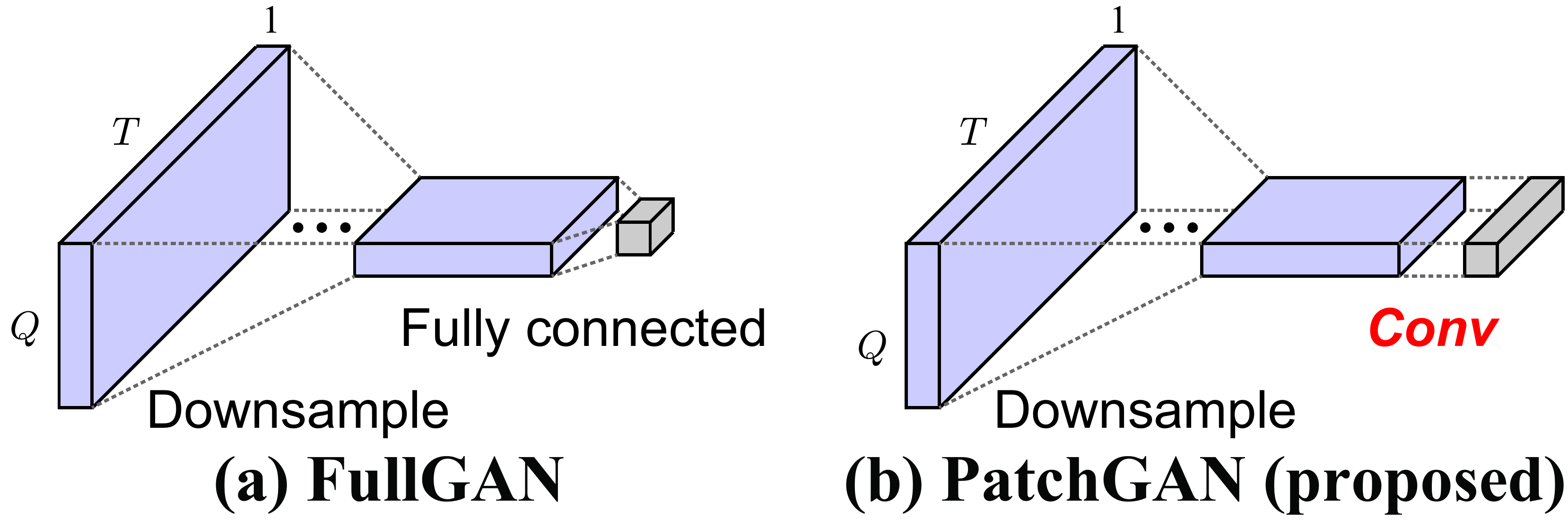}}
  \vspace{-3.5mm}
  \caption{Comparison of discriminator network architectures}
  \vspace{-5mm}
  \label{fig:discriminator}
\end{figure}

In previous GAN-based speech processing models~\cite{TKanekoICASSP2017,TKanekoIS2017a,TKanekoIS2017b,TKanekoArXiv2017}, FullGAN (Fig.~\ref{fig:discriminator}(a)) has been extensively used. However, recent studies in computer vision~\cite{PIsolaCVPR2017,WShiCVPR2016} indicate that the wide-range receptive fields of the discriminator require more parameters, which causes difficulty in training. Inspired by this, we replace FullGAN with \textit{PatchGAN}~\cite{CLiECCV2016,PIsolaCVPR2017,WShiCVPR2016} (Fig.~\ref{fig:discriminator}(b)), which uses convolution at the last layer and determines the realness on the basis of the patch. We experimentally examine its effect for non-parallel VC in Section~\ref{subsec:exp_obj}.

\section{Experiments}
\vspace{-1mm}
\label{sec:exp}
\subsection{Experimental Conditions}
\vspace{-1mm}
\label{subsec:exp_cond}
{\bf Dataset:}
We evaluated our method on the {\bf Spoke} (i.e., {\bf non-parallel VC}) task of the VCC 2018~\cite{VCC2018}, which includes recordings of professional US English speakers. We selected a subset of speakers so as to cover all inter-gender and intra-gender conversions: VCC2SF3 (\bfsl{SF}), VCC2SM3 (\bfsl{SM}), VCC2TF1 (\bfsl{TF}), and VCC2TM1 (\bfsl{TM}), where S, T, F, and M indicate source, target, female, and male, respectively. In the following, we use the abbreviations in the parenthesis (e.g., \bfsl{SF}). Combinations of 2 sources (\bfsl{SF} or \bfsl{SM}) $\times$ 2 targets (\bfsl{TF} or \bfsl{TM}) were used for evaluation. Each speaker has sets of 81 (about 5 minutes; {\bf relatively few} for VC) and 35 sentences for training and evaluation, respectively. In the Spoke task, the source and target speakers have a {\bf different} set of sentences ({\bf no overlap}) so as to evaluate in a non-parallel setting. The recordings were downsampled to 22.05 kHz for this challenge. We extracted 34 Mel-cepstral coefficients (MCEPs), logarithmic fundamental frequency ($\log F_0$), and aperiodicities (APs) every 5 ms by using the WORLD analyzer~\cite{MMoriseIEICE2016}.

{\bf Conversion process:}
The proposed method was used to convert MCEPs ($Q = 34 + 1$ dimensions including $0$th coefficient).\footnote{For reference, the converted speech samples, in which the proposed method was used to convert all acoustic features (namely, MCEPs, band APs, continuous $\log F_0$, and voice/unvoice indicator), are provided at \url{http://www.kecl.ntt.co.jp/people/kaneko.takuhiro/projects/cyclegan-vc2/index.html}. Even in this challenging setting, CycleGAN-VC2 works reasonably well.} The objective of these experiments was to analyze the quality of the converted MCEPs; therefore, for the other parts, we used typical methods similar to the baseline of the VCC 2018~\cite{VCC2018}. Specifically, in inter-gender conversion, a vocoder-based VC method was used. $F_0$ was converted by using logarithm Gaussian normalized transformation~\cite{KLiuFSKD2007}, APs were directly used without modification, and the WORLD vocoder~\cite{MMoriseIEICE2016} was used to synthesize speech. In intra-gender conversion, we used a vocoder-free VC method~\cite{KKobayashiSLT2016}. More precisely, we calculated differential MCEPs by taking the difference between the source and converted MCEPs. For a similar reason, we did not use any postfilter~\cite{TKanekoICASSP2017,TKanekoIS2017a,KTanakaSLT2018} or powerful vocoder such as the WaveNet vocoder~\cite{OAaronArXiv2016,ATamamoriIS2017}. Incorporating them is one possible direction of future work.

\begin{figure*}[t]
  \centerline{\includegraphics[width=\textwidth]{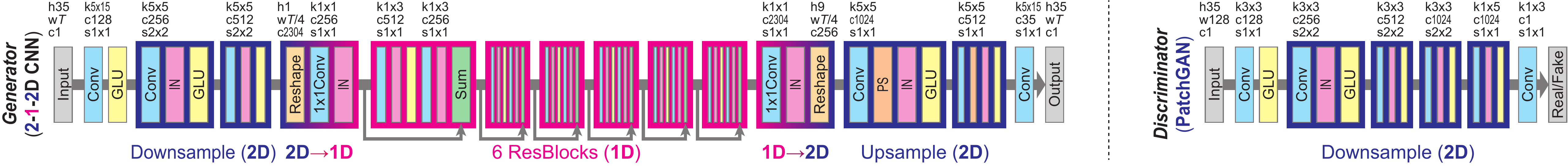}}
  \vspace{-5mm}
  \caption{Network architectures of generator and discriminator. In input, output, and reshape layers, {\tt h}, {\tt w}, and {\tt c} denote height, width, and number of channels, respectively. In each convolution layer, {\tt k}, {\tt c}, and {\tt s} denote kernel size, number of channels, and stride size, respectively. IN, GLU, and PS indicate instance normalization~\cite{DUlyanovArXiv2016}, gated linear unit~\cite{YDauphinICML2017}, and pixel shuffler~\cite{WShiCVPR2016}, respectively. Since the generator is fully convolutional~\cite{JLongCVPR2015}, it can take input of arbitrary length $T$.}
  \vspace{-6mm}
  \label{fig:network}
\end{figure*}

{\bf Training details:}
The implementation was almost the same as that of CycleGAN-VC except that the improved techniques were incorporated. The details of the network architectures are given in Fig.~\ref{fig:network}. For a pre-process, we normalized the source and target MCEPs to zero-mean and unit-variance by using the statistics of the training sets. To stabilize training, we used a least squares GAN (LSGAN)~\cite{XMaoICCV2017}. To increase the randomness of training data, we randomly cropped a segment (128 frames) from a randomly selected sentence instead of using an overall sentence directly. We used the Adam optimizer~\cite{DPKingmaICLR2015} with a batch size of 1. We trained the networks for $2 \times 10^5$ iterations with learning rates of 0.0002 for the generator and 0.0001 for the discriminator and with momentum term $\beta_1$ of 0.5. We set $\lambda_{cyc} = 10$ and $\lambda_{id} = 5$. We used ${\cal L}_{id}$ only for the first $10^4$ iterations to guide the learning direction. {\bf Note that we did not use any extra data, modules, or time alignment procedures for training}.

\subsection{Objective Evaluation}
\vspace{-1mm}
\label{subsec:exp_obj}

\begin{table}[t]  
  \centering
  \vspace{-3mm}
  \caption{Comparison of MCD [dB]}
  \label{tab:mcd}
  \scriptsize{
    \begin{tabular*}{1.01\columnwidth}{c|c|c|c|cc|cc}
      \!\!\!No.\!\!\!& \multicolumn{3}{c|}{\bf Method}
      & \multicolumn{2}{c|}{\bf Intra-gender}
      & \multicolumn{2}{c}{\bf Inter-gender} \\      
      \hline
      & \multicolumn{3}{c|}{CycleGAN-VC2}
      & \multirow{2}{*}{SF-TF}
      & \multirow{2}{*}{SM-TM}
      & \multirow{2}{*}{SM-TF}
      & \multirow{2}{*}{SF-TM} \\
      \cline{2-4}
      & Adv. & $G$ & $D$
      & \multicolumn{2}{c|}{ } \\
      \hline
      \!1\!& \!1Step\! & \!2-1-2D\! & \!Patch\!
      & \!6.86$\pm$.04\! & \!6.32$\pm$.06\!
      & \!7.36$\pm$.04\! & \!6.28$\pm$.04\! \\
      \!2\!& \!2Step\! & 1D & \!Patch\!
      & \!6.86$\pm$.04\! & \!6.73$\pm$.08\! 
      & \!7.77$\pm$.07\! & \!6.41$\pm$.01\! \\
      \!3\!& \!2Step\! & 2D & \!Patch\!
      & \!7.01$\pm$.07\! & \!6.63$\pm$.03\!
      & \!7.63$\pm$.03\! & \!6.73$\pm$.04\! \\
      \!4\!& \!2Step\! & \!2-1-2D\! & \!Full\!
      & \!7.01$\pm$.07\! & \!6.45$\pm$.05\! 
      & \!7.41$\pm$.04\! & \!6.51$\pm$.02\! \\
      \!5\!& \!{\bf 2Step}\! & \!{\bf 2-1-2D}\! & \!{\bf Patch}\!
      & \!{\bf 6.83$\pm$.01}\! & \!{\bf 6.31$\pm$.03}\! 
      & \!{\bf 7.22$\pm$.05}\! & \!{\bf 6.26$\pm$.03}\! \\
      \hline
      \!6\!& \multicolumn{3}{c|}{CycleGAN-VC~\cite{TKanekoArXiv2017}}
      & \!7.37$\pm$.03\! & \!6.68$\pm$.07\! 
      & \!7.68$\pm$.05\! & \!6.51$\pm$.05\! \\
      \!7\!& \multicolumn{3}{c|}{\!\!\!Frame-based CycleGAN~\cite{FFangICASSP2018}\!\!\!}
      & \!8.85$\pm$.07\! & \!7.27$\pm$.11\!
      & \!8.86$\pm$.27\! & \!8.51$\pm$.36\! \\
    \end{tabular*}
  }
  \vspace{-3mm}
\end{table}

\begin{table}[t]  
  \centering
  \vspace{-4mm}
  \caption{Comparison of MSD [dB]}
  \label{tab:msd}
  \scriptsize{
    \begin{tabular*}{1.01\columnwidth}{c|c|c|c|cc|cc}
      \!\!\!No.\!\!\!& \multicolumn{3}{c|}{\bf Method}
      & \multicolumn{2}{c|}{\bf Intra-gender}
      & \multicolumn{2}{c}{\bf Inter-gender} \\      
      \hline      
      & \multicolumn{3}{c|}{CycleGAN-VC2}
      & \multirow{2}{*}{SF-TF}
      & \multirow{2}{*}{SM-TM}
      & \multirow{2}{*}{SM-TF}
      & \multirow{2}{*}{SF-TM} \\
      \cline{2-4}
      & Adv. & $G$ & $D$
      & \multicolumn{2}{c|}{ } \\
      \hline
      \!1\!& \!1Step\! & \!2-1-2D\! & \!Patch
      & \!1.60$\pm$.02\! & \!1.63$\pm$.05\!
      & \!1.54$\pm$.03\! & \!1.56$\pm$.04\! \\
      \!2\!& \!2Step\! & \!1D\! & \!Patch
      & \!3.31$\pm$.36\! & \!4.26$\pm$.37\!
      & \!2.04$\pm$.21\! & \!5.03$\pm$.32\! \\
      \!3\!& \!2Step\! & \!2D\! & \!Patch
      & \!1.57$\pm$.07\! & \!1.54$\pm$.01\!
      & \!1.46$\pm$.03\! & \!1.66$\pm$.07\! \\
      \!4\!& \!2Step\! & \!2-1-2D\! & \!Full
      & \!1.52$\pm$.02\! & \!1.56$\pm$.04\!
      & \!1.47$\pm$.01\! & \!1.67$\pm$.06\! \\
      \!5\!& \!{\bf 2Step}\! & \!{\bf 2-1-2D}\! & \!{\bf Patch}
      & \!{\bf 1.49$\pm$.01}\! & \!{\bf 1.53$\pm$.02}\!
      & \!{\bf 1.45$\pm$.00}\! & \!{\bf 1.52$\pm$.01}\! \\
      \hline
      \!6\!& \multicolumn{3}{c|}{CycleGAN-VC~\cite{TKanekoArXiv2017}}
      & \!2.42$\pm$.08\! & \!2.66$\pm$.08\!
      & \!2.21$\pm$.13\! & \!2.65$\pm$.15\! \\
      \!7\!& \multicolumn{3}{c|}{\!\!\!Frame-based CycleGAN~\cite{FFangICASSP2018}\!\!\!}
      & \!3.78$\pm$.26\! & \!2.77$\pm$.10\!
      & \!3.32$\pm$.06\! & \!3.61$\pm$.15\!  \\
    \end{tabular*}
  }
  \vspace{-6mm}
\end{table}

As discussed in previous studies~\cite{TTodaTASLP2007,TKanekoICASSP2017}, it is fairly complex to design a single metric that can assess the quality of converted MCEPs comprehensively. Alternatively, we used two metrics to assess the local and global structures. To measure global structural differences, we used the Mel-cepstral distortion (MCD), which measures the distance between the target and converted MCEP sequences. To measure the local structural differences, we used the modulation spectra distance (MSD), which is defined as the root mean square error between the target and converted logarithmic modulation spectra of MCEPs averaged over all MCEP dimensions and modulation frequencies. For both metrics, smaller values indicate that target and converted MCEPs are more similar.

We list the MCD and MSD in Tables~\ref{tab:mcd} and \ref{tab:msd}, respectively. To eliminate the effect of initialization, we report the average and standard deviation scores over three random initializations. To analyze the effect of each technique, we conducted ablation studies on \bfsl{CycleGAN-VC2} (no. 5 is the full model). We also compared \bfsl{CycleGAN-VC2} with two state-of-the-art methods: \bfsl{CycleGAN-VC}~\cite{TKanekoArXiv2017} and \bfsl{frame-based CycleGAN}~\cite{FFangICASSP2018} (our reimplementation; we additionally used ${\cal L}_{id}$ for stabilizing training). The comparison of one-step and two-step adversarial losses (nos. 1, 5) indicates that this technique is particularly effective for improving MSD. The comparisons of generator (nos. 2, 3, 5) and discriminator (nos. 4, 5) network architectures indicate that they contribute to improving both MCD and MSD. Finally, the comparison to the baselines (nos. 5, 6, 7) verifies that by incorporating the three proposed techniques, we achieve state-of-the-art performance in terms of MCD and MSD for every speaker pair.

\vspace{-3.8mm}
\subsection{Subjective Evaluation}
\begin{figure}[t]
  \centerline{\includegraphics[width=\columnwidth]{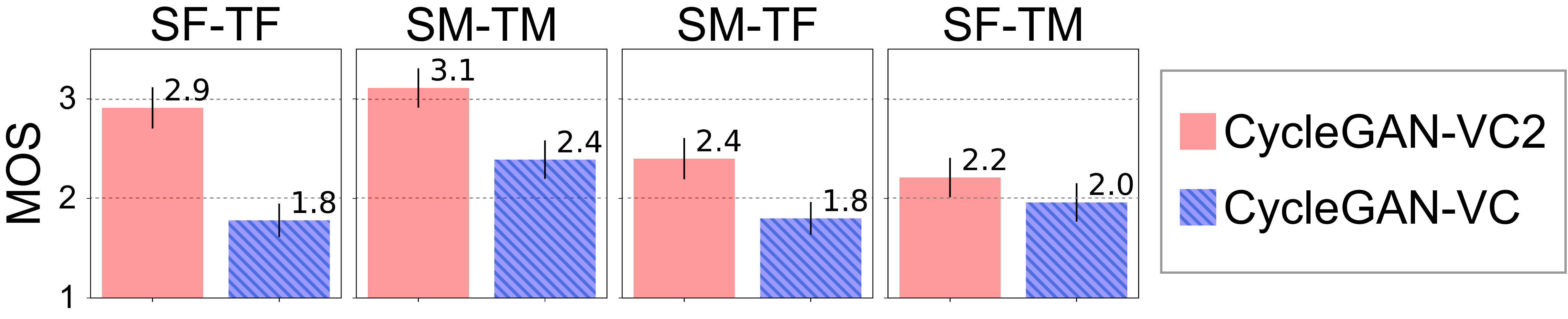}}
  \vspace{-4mm}
  \caption{MOS for naturalness with 95\% confidence intervals}
  \vspace{-1mm}
  \label{fig:mos}
\end{figure}

\begin{figure}[t]
  \centerline{\includegraphics[width=0.94\columnwidth]{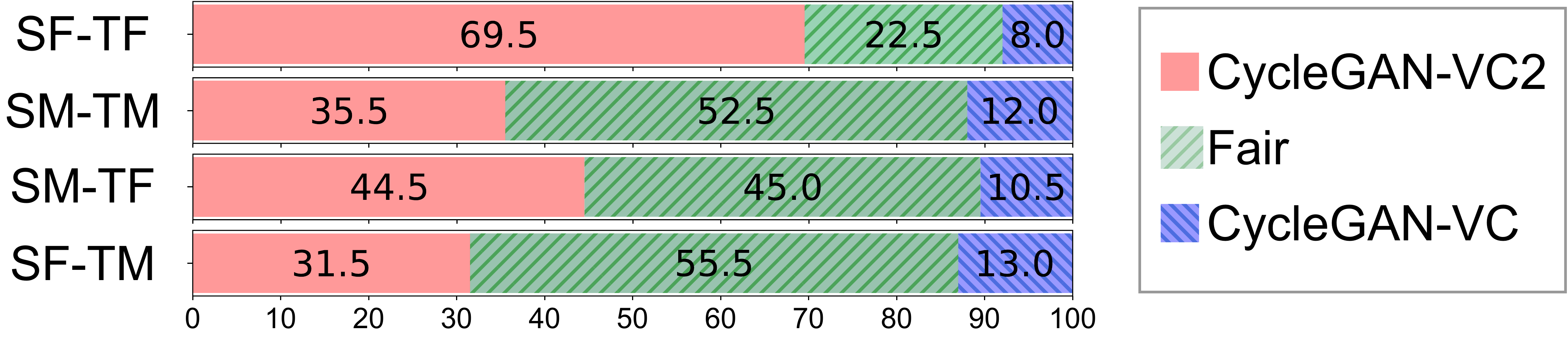}}
  \vspace{-4mm}
  \caption{Average preference score (\%) on speaker similarity}
  \vspace{-5mm}
  \label{fig:xab}
\end{figure}

\vspace{-1mm}
\label{subsec:exp_subj}
We conducted listening tests to evaluate the quality of converted speech. \bfsl{CycleGAN-VC}~\cite{TKanekoArXiv2017} was used as the baseline. To measure naturalness, we conducted a mean opinion score (MOS) test (5: excellent to 1: bad), in which we included the target speech as a reference (MOS for \bfsl{TF} and \bfsl{TM} are 4.8). Ten sentences were randomly selected from the evaluation sets. To measure speaker similarity, we conducted an XAB test, where ``A'' and ``B'' were speech converted by the baseline and proposed methods, and ``X'' was target speech. We selected ten sentence pairs randomly from the evaluation sets and presented all pairs in both orders (AB and BA) to eliminate bias in the order of stimuli. For each sentence pair, the listeners were asked to select their preferred one (``A'' or ``B'') or to opt for ``Fair.'' Ten listeners participated in these listening tests. Figs.~\ref{fig:mos} and \ref{fig:xab} show the MOS for naturalness and the preference scores for speaker similarity, respectively. These results confirm that \bfsl{CycleGAN-VC2} outperforms \bfsl{CycleGAN-VC} in terms of both naturalness and similarity for every speaker pair. Particularly, \bfsl{CycleGAN-VC} is difficult to apply to a vocoder-free VC framework~\cite{KKobayashiSLT2016} (used in \bfsl{SF-TF} and \bfsl{SM-TM}), as this framework is sensitive to conversion error due to the usage of differential MCEPs. However, the MOS indicates that \bfsl{CycleGAN-VC2} works relatively well in such a difficult setting.

\section{Conclusions}
\label{sec:conc}
To advance the research on non-parallel VC, we have proposed CycleGAN-VC2, which is an improved version of CycleGAN-VC incorporating three new techniques: an improved objective (two-step adversarial losses), improved generator (2-1-2D CNN), and improved discriminator (PatchGAN). The experimental results demonstrate that CycleGAN-VC2 outperforms CycleGAN-VC in both objective and subjective measures for every speaker pair. Our proposed techniques are not limited to one-to-one VC, and adapting them to other settings (e.g., multi-domain VC~\cite{HKameokaSLT2018}) and other applications~\cite{AKainSC2007,KNakamuraSC2012,TTodaTASLP2012,ZInanogluSC2009,TKanekoIS2017b} remains an interesting future direction.

\noindent
{\bf Acknowledgements:}
This work was supported by JSPS KAKENHI 17H01763.

\clearpage
\bibliographystyle{IEEEbib}
\scriptsize{\bibliography{refs}}

\end{document}